\documentclass[aps,prb,twocolumn,superscriptaddress]{revtex4}
\usepackage{graphicx}
\usepackage{amssymb}
\usepackage{amsmath}
\usepackage{latexsym,bm,array,amsfonts,multirow}
\usepackage{epsfig}
\usepackage{color}
\usepackage{ulem}

\makeatletter
\newsavebox{\@brx}
\newcommand{\llangle}[1][]{\savebox{\@brx}{\(\m@th{#1\langle}\)}%
	\mathopen{\copy\@brx\kern-0.5\wd\@brx\usebox{\@brx}}}
\newcommand{\rrangle}[1][]{\savebox{\@brx}{\(\m@th{#1\rangle}\)}%
	\mathclose{\copy\@brx\kern-0.5\wd\@brx\usebox{\@brx}}}
\makeatother

\begin{document}

\title{Topological phase transition driven by   Hatsugai-Kohmoto interaction on the Kagome lattice}
\author{Zhuo-Xuan Yang}
\affiliation{School of Physics, Hangzhou Normal University, Hangzhou, Zhejiang 311121, China}
\author{Jiangfan Wang}
\email[]{jfwang@hznu.edu.cn}
\affiliation{School of Physics, Hangzhou Normal University, Hangzhou, Zhejiang 311121, China}
\date{\today}

\begin{abstract}
The interplay between band topology and strong correlations is central to modern condensed matter physics, but exact solutions are rare. Here, we present an exactly solvable model on the Kagome lattice by combining a Kane-Mele-type spin-orbit coupling with the Hatsugai-Kohmoto interaction. At $1/3$ filling, we uncover a continuous topological quantum phase transition driven by electron interaction. A weakly correlated $\mathbb{Z}_2$ topological insulator gives way to a strongly correlated insulator that, while $\mathbb{Z}_2$-trivial, hosts a nontrivial spin Chern number $C_s=2$. The transition exhibits critical scaling consistent with the universality class of two-dimensional Dirac fermions. At half-filling, the same model yields a non-Fermi-liquid to Mott-insulator transition, demonstrating that correlation-driven topological and Mott transitions can be unified within a single solvable framework. Our results establish the Kagome Hatsugai-Kohmoto model as a valuable benchmark for interacting topological systems.     
\end{abstract}
\maketitle

\section{Introduction}
The interplay between band-structure topology and electron correlation remains a central topic in condensed matter physics \cite{Rachel2018review,Assaad2013Review}. On one hand, the discovery of topological (Chern) insulators and topological superconductors has attracted great attention due to their symmetry-protected edge or surface states and potential device applications \cite{Hasan2010RMP,Zhang2011RMP,KaneMele2005}.
On the other hand, electron correlation naturally exists in quantum materials, and is the cause for various emergent phenomena such as the high-temperature superconductivity \cite{Lee_Nagaosa_Wen2006RMP} and metal-Mott insulator transitions\cite{Imada1998RMP}. The combination of these two themes leads to fruitful discoveries in the past decades, ranging from weakly correlated topological insulator to strongly correlated topological Kondo insulator \cite{Coleman2010PRL,Dai2013SmB6} and topological Mott insulator \cite{Raghu2008PRL,Pesin2010NatPhy,Wagner2023NatCom}. Theoretically, there are two major technical challenges: First, the electron correlation is often modeled by the Hubbard interaction, which is notoriously difficult to deal with and inevitably requires various analytical or numerical approximations \cite{Hichey2015,Rachel2010PRB_KM,Assaad2011PRL_KM,CJWu2011PRB_KM,Budich2012PRB,Budich2013PRB_BHZ,Ruegg2013PRB,Amaricci2015PRL,Raghu2008PRL,Zeng2017PRB}. Second, the mathematical tools developed for calculating topological invariants in noninteracting systems are not directly applicable for strongly correlated systems where single-particle pictures break down\cite{Pasqua2025gCut,Phillps2023PRL}.

Recently, the Hatsugai-Kohmoto (HK) model with local-in-momentum-space interactions have gained growing attention due to their exact solvability and close relations to the Hubbard model and strongly correlated physics \cite{HK1992,Phillips2022NatPhys,Phillips2020NatPhys,Phillips2023fixPoint,Phillips2024Hubbard,Zhong2025review,Zhong2022PRB,YLi2022BCS,Wang2023Kondo,Wang2024HeavyFermion}. Although with a seemingly unrealistic infinite-range interaction, the HK model has been demonstrated to represent a stable quartic fixed point of the Hubbard model around half-filling \cite{Phillips2022NatPhys,Phillips2023fixPoint}, and recovers many important features of the Hubbard model once the momentum mixing is partially re-introduced \cite{Phillips2024Hubbard}.  It is therefore suitable for studying correlation effects on band topology by combining HK interaction with relevant Hamiltonians such as the Haldane model \cite{Phillips2023PRR}, the Kane-Mele model \cite{Phillips2024PRB,Desort2025PRB,Bradlyn2023PRB,Skolimowski2025PRB} and the Bernevig-Hughes-Zhang model \cite{Phillips2023NatCom}. Previous studies have led to interesting results such as topological Mott insulators with odd-integer fillings \cite{Phillips2023PRR,Phillips2023NatCom,Jablonowski2023Anderson} and topological phase transitions without a gap closing \cite{Phillips2024PRB,Jablonowski2023Anderson,Brzezicki2023PRB}.

In this paper, we investigate the interplay between band topology and electron correlation via a topological HK model on the Kagome lattice. The Kagome lattice has attracted significant interest in recent years due to the existence of flat band and Dirac fermions \cite{JXYin2022Review,CJWu2008PRB}, which can become topologically nontrivial once spin-orbit coupling \cite{Wen2011PRL,Guo2009,GLiu2010PRA,Xu2015PRL,Bolens2019PRB,Kang2020NatCom} or electron interaction \cite{JunWen2010PRB,WZhu2016PRL} is included. Moreover, the multi-band structure of Kagome lattice allows us to study topological and Mott transitions at separate fillings. By exactly solving this model and calculating both the $\mathbb{Z}_2$ invariant and spin Chern number, we show a continuous topological quantum phase transition driven by interaction at exactly 1/3 filling. The critical behavior is consistent with the topological phase transition of a two-dimensional Dirac Hamiltonian. At half-filling, we found a transition from a non-Fermi liquid metal to a Mott insulator with internal spin entanglement between different sublattices. Our work provides an exactly solvable platform on a multi-band frustrated lattice that unifies correlation-driven topological quantum phase transitions and Mott metal-insulator transitions within a single framework.

\section{Model and Methods}

\begin{figure}[t]
	\begin{center}
		\includegraphics[width=8cm]{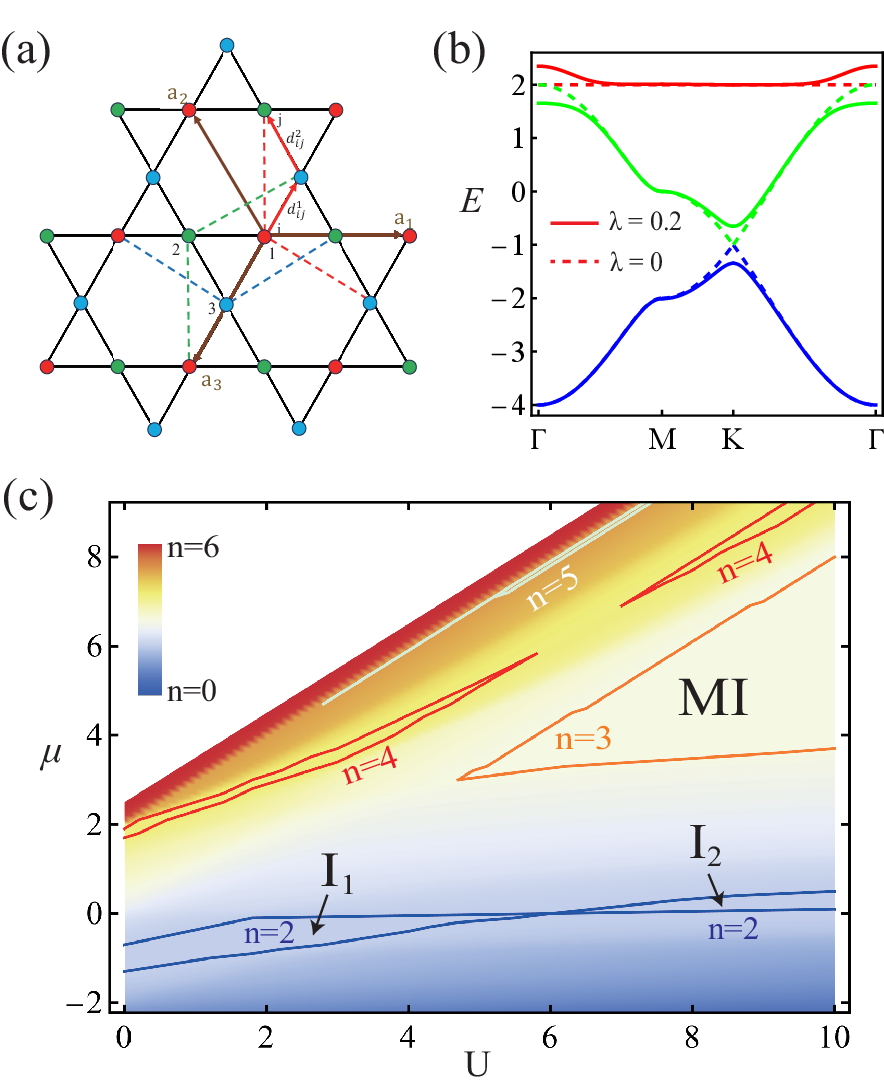}
	\end{center}
	\caption{(a) An illustration of the unit vectors and hopping parameters on the Kagome lattice. (b) The free electron band structure of Hamiltonian Eq. (\ref{eq:H0}) along the high symmetry points with spin-orbit coupling $\lambda=0$ (dashed lines) and $0.2$ (solid lines). (c) The ground state phase diagram in terms of $\mu$ and HK interaction $U$ from a color-coded plot of electron occupation $n$. MI denotes the Mott insulator, while $I_1$  and $I_2$ denote two topologically distinct insulators.}
	\label{fig1}
\end{figure}

\textit{Model.---}We start from the following non-interacting Hamiltonian defined on the Kagome lattice \cite{Guo2009}:
\begin{equation}
	H_0=\sum_{ ij\sigma}(t_{ij}-\mu\delta_{ij})c_{i\sigma}^{\dagger}c_{j\sigma}+\frac{i\lambda}{2}\sum_{\llangle ij\rrangle\alpha\beta}\nu_{ij} c_{i\alpha}^\dagger \sigma_{\alpha\beta}^z c_{j\beta}, \label{eq:H0}
\end{equation}
where the first term describes free electrons with nearest-neighbor hopping amplitude $t_{ij}=-t$ and chemical potential $\mu$, the second term describes an intrinsic spin-orbit coupling with strength $\lambda$ for next-nearest-neighbor hoppings \cite{Guo2009,KaneMele2005}. $\nu_{ij}=\pm 1$ depends on the orientation of the sites, formally defined as $\nu_{ij}=\frac{8}{\sqrt{3}}(\mathbf{d}_{ij}^1\times \mathbf{d}_{ij}^2)_z$, with $\mathbf{d}_{ij}^{1,2}$ being the lattice vectors connecting the sites $i$ and $j$ (see Fig. (\ref{fig1}a)). For simplicity, the Rashba spin-orbit interaction is not considered here, so that the U(1) spin rotational symmetry is preserved.

Upon Fourier transform, Eq. (\ref{eq:H0}) can be written in the momentum space as:
\begin{eqnarray}
	H_0&=&\sum_{\mathbf{k}l \sigma}\left(i\lambda\tilde{\sigma}\cos (k_{l+1}-k_{l+2})-2t\cos k_l\right)c_{\mathbf{k}l\sigma}^\dagger c_{\mathbf{k},l+1,\sigma}\notag\\
	&&+H.c.-\mu\sum_{\mathbf{k}l\sigma}n_{\mathbf{k}l\sigma},\label{eq:H0-1}
\end{eqnarray}
where $\tilde{\sigma}=1(-1)$ for spin up (down), $l\in\{1,2,3\}$ denotes the three sublattices with unit vectors $\mathbf{a}_1=(1,0)$, $\mathbf{a}_2=(-1/2,\sqrt{3}/2)$, and $\mathbf{a}_3=-\mathbf{a}_1-\mathbf{a}_2=(-1/2,-\sqrt{3}/2)$. We have denoted $k_l=\mathbf{k}\cdot \mathbf{a}_l/2$ and the density operator $n_{\mathbf{k}l\sigma}=c_{\mathbf{k}l\sigma}^\dagger c_{\mathbf{k}l\sigma}$. Diagonalizing Eq. (\ref{eq:H0-1}) gives rise to three bands, whose dispersion relations are shown in Fig. (\ref{fig1}b) for $\lambda=0$ and $\lambda=0.2$. A finite spin-orbit coupling opens a gap both between the top flat band and the middle dispersive band, and between the middle and bottom dispersive bands. As has been studied in Ref.\cite{Guo2009}, both gaps are  topologically nontrivial with a $\mathbb{Z}_2$ invariant $\nu=1$.

To study the electron correlation effects, we introduce the following form of HK interaction,
\begin{eqnarray}
	H_U&=&U\sum_{\mathbf{k}l}n_{\mathbf{k}l\uparrow}n_{\mathbf{k}l\downarrow},\label{eq:HU}
\end{eqnarray}
which transforms to an infinite-range two-particle scattering term in the coordinate space that preserves the center of mass on each sublattice \cite{HK1992}. Note that Eq. (\ref{eq:HU}) mixes electrons on different original bands shown in Fig. (\ref{fig1}b), which differs from previous ones formulated directly in band basis \cite{Phillips2023PRR,Phillips2023NatCom}. Here, we do not consider the inter-sublattice repulsion, which appears to be necessary for ``gap-not-closing" topological phase transitions in the Kane-Mele-HK model\cite{Phillips2024PRB}. Our calculations show that the intra-sublattice repulsion given above is already enough to cause interesting topological phase transitions. The combined Hamiltonian  $H=H_0+H_U=\sum_{\mathbf{k}}H_\mathbf{k}$ can be exactly solved since $[H_\mathbf{k},H]=0$ for all momenta. Each $H_\mathbf{k}$ can be viewed as a three-site Hubbard molecule with a Hilbert-space dimension $4^3=64$, which can be further divided into 16 subspaces using the conserved quantity $n_{\mathbf{k}\sigma}=\sum_l n_{\mathbf{k}l\sigma }$, hence is easily diagonalized.

\textit{Topological invariant.---}Since our model preserves the time-reversal symmetry, the spatial-inversion symmetry and the U(1)
spin-rotational symmetry, we can use both the $\mathbb{Z}_2$ invariant and the spin Chern number to classify different topological states. Since the HK interaction now mixes different bands, the Chern number cannot be directly obtained from the original bands. Instead, we use the eigenstates of zero-frequency Green's function obtained exactly to calculate the topological invariants \cite{SCZhang2012,ZYLu2016a,ZYLu2016b}. The inverse of zero-frequency Green's function is also known as the ``topological Hamiltonian" for correlated systems \cite{BHYan2013}, whose eigenstates play a similar role as the Bloch states in non-interacting systems. The $\mathbb{Z}_2$ invariant can be calculated as 
\begin{eqnarray}
	(-1)^\nu=\prod_{\mathbf{k}\in \text{TRIM}} \prod_{\mu_\eta>0}\langle \phi_{\sigma\eta}(\mathbf{k})|P_{\mathbf{k}}|\phi_{\sigma\eta}(\mathbf{k}) \rangle, \label{eq:Z2}
\end{eqnarray}
where TRIM represents the time-reversal invariant momentum points, $\mathbf{\Gamma}_i=(0,0),(0,2\pi/\sqrt{3}),(\pi, \pi/\sqrt{3})$, $(-\pi,\pi/\sqrt{3})$, and $P_{\mathbf{k}}$ is the parity operator.  $\mu_\eta(\mathbf{k})$ and $\left|\phi_{\sigma\eta}(\mathbf{k})\right\rangle$ (with $\eta=1,2,3$) are the eigenvalues and eigenstates of the $3\times 3$ Green's function matrix for each spin component $\sigma$,
\begin{equation}
 G_\sigma(i\omega=0,\mathbf{k})\left|\phi_{\sigma\eta}(\mathbf{k})\right\rangle=\mu_\eta(\mathbf{k})\left|\phi_{\sigma\eta}(\mathbf{k})\right\rangle.
\end{equation} 
Note that we only need eigenstates with positive $\mu_\eta$, corresponding to ``occupied bands" of the topological Hamiltonian $h_t(\mathbf{k})=\text{diag}(-G^{-1}_\uparrow (0,\mathbf{k}),-G^{-1}_\downarrow(0,\mathbf{k}))$.

Equation (\ref{eq:Z2}) is not directly applicable if any TRIM point has degenerate ground states, as encountered in our calculations for $1/3$ filling at large $U$. In that case, we adopt an alternative definition of the $\mathbb{Z}_2$ invariant, namely, the parity of the number of zeros of the Pfaffian in half of the Brillouin zone (BZ) \cite{KaneMele2005}. The Pfaffian is constructed from the antisymmetric matrix of overlaps between occupied eigenstates of the topological Hamiltonian $h_t(\mathbf{k})$ and their time-reversed partners. For the $1/3$ filling case of our model, it reduces to
\begin{equation}
	P(\mathbf{k}) =\langle \phi_{\uparrow \eta}(\mathbf k)| \phi_{\downarrow \eta}(\mathbf{-k})\rangle, \label{eq:Pf}
\end{equation}  
where $|\phi_{\sigma \eta}(\mathbf{k}) \rangle$ is the eigenstate of $G_\sigma(0,\mathbf{k})$ with positive eigenvalue $\mu_\eta (\mathbf{k})=\mu_\eta (\mathbf{-k})>0$. At the TRIM points, one always has $|P(\mathbf k)|=1$, regardless of which degenerate ground state is chosen to compute $G_\sigma(0,\mathbf{k})$. For all other momenta, the ground state is unique, and Eq. (\ref{eq:Pf}) remains well-defined.

In the presence of U(1) spin symmetry, the system's topological property is also conveniently characterized by the spin Chern number \cite{Sheng2006,Prodan2009}, defined as
\begin{eqnarray}
	C_s&=&\frac{C_\uparrow-C_\downarrow}{2}=C_\uparrow \notag \\
	&=&\frac{1}{2\pi}\int_{\mathbf{k}\in\text{BZ}}d^2\mathbf{k}\left( \partial_{k_x}\mathcal{A}_y(\mathbf{k})-\partial_{k_y}\mathcal{A}_x(\mathbf{k})\right),\label{eq:Cs}
\end{eqnarray}
where $\mathcal{A}_{i}(\mathbf{k})=-i\sum_{\mu_\eta>0}\langle \phi_{\sigma \eta}(\mathbf{k})|\partial_{k_i}|\phi_{\sigma\eta}(\mathbf{k}) \rangle$ is the Berry vector potential obtained from the eigenstates of zero-frequency Green's function, and $C_\uparrow$ ($C_\downarrow$) denotes the Chern number for the spin up (down) component. The spin Chern number is unaffected by the ground-state degeneracy at isolated momentum points, since these points constitute a measure-zero set of the two-dimensional momentum integral in Eq. (\ref{eq:Cs}), whose contribution vanishes in the thermodynamic limit.

\section{Results}

\subsection{Phase diagram}

The ground state phase diagram in terms of $\mu$ and $U$ is shown in Fig. (\ref{fig1}c) with a color-coded plot of the electron occupation number  $n=\mathcal{N}_s^{-1}\sum_{\mathbf{k}l\sigma}n_{\mathbf{k}l\sigma}$ for $\lambda=0.2$. One can find several insulating phases characterized by different integer values of $n$. For $n=2$ (corresponding to $1/3$ filling), there are two different insulating phases separated by a quantum phase transition at a critical point $U_c\approx 6$. As we will show later, both the $\mathbb{Z}_2$ invariant (calculated from the Pfaffian) and the spin Chern number jumps at $U_c$, revealing a topological quantum phase transition driven by electron correlation. Similar topological phase transition also occurs for $n=4$ ($2/3$ filling), with roughly the same critical interaction $U_c\approx 6$. For $n=3$ (half filling), one finds a phase transition between a non-Fermi-liquid metal at small $U$ and a Mott insulating phase at large $U$. Similar Mott transitions also occur at other odd-integer fillings $n=5$ and $n=1$. Note that the $n=1$ Mott transition occurs at very large  $U$ and is not shown in Fig. (\ref{fig1}c).

\begin{figure*}[t]
	\begin{center}
		\includegraphics[width=13cm]{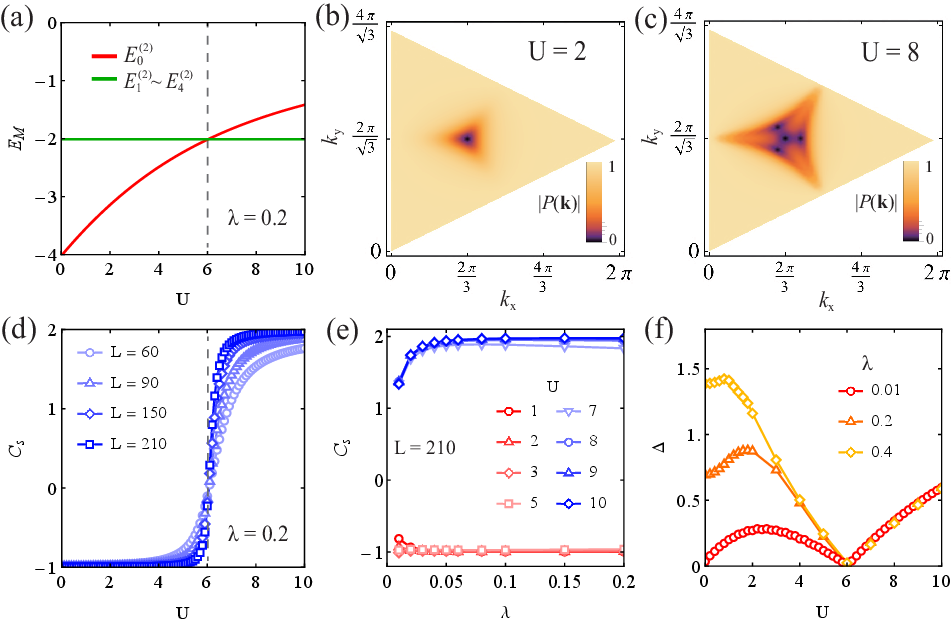}
	\end{center}
	\caption{Topological phase transition at $1/3$ filling. (a) Evolution of the five lowest eigenvalues of $H_{\mathbf{k}=M}^{(2)}$ for $\lambda=0.2$, showing a level crossing at $U_c\approx 6$. (b)(c) Density plots of the absolute value of Pfaffian $|P(\mathbf{k})|$ within half of the Brillouin zone,  showing one and four zeros (black points) for $U=2$ and $U=8$, respectively. (d) The spin Chern number $C_s$ as a function of $U$ for $\lambda=0.2$,  with different system sizes $L=60 \sim 210$. (e) $C_s$ as functions of $\lambda$ for different values of $U$. (f) Evolution of the single-particle gap ($\Delta$) with increasing $U$ for different $\lambda$.}
	\label{fig2}
\end{figure*}

\subsection{Topological phase transition at $1/3$ filling}

We first focus on the interaction-driven topological phase transition at $1/3$ filling. In this case, the calculations are simplified since every momentum point within the first BZ is occupied by two electrons at zero temperature, so that only the $n_{\bf k}=2$ block of $H_{\bf k}$ ($H_{\bf k}^{(2)}$) is required to obtain the ground state, and the $n_{\bf k}=1,3$ blocks ($H_{\bf k}^{(1)}$ and $H_{\bf k}^{(3)}$) for the single-particle excited states. Using the Lehmann representation, the zero-temperature Green's function at $\omega=0$ can be written as:
\begin{eqnarray}
	G_{\sigma}^{ll'}(0,\mathbf{k})&=&\sum_{m=0}^{19}\frac{\langle \psi_0^{(2)}| c_{\mathbf{k}l\sigma}|\psi_m^{(3)} \rangle \langle \psi_m^{(3)}| c_{\mathbf{k}l'\sigma}^\dagger|\psi_0^{(2)} \rangle}{E_{0,\mathbf{k}}^{(2)}-E_{m,\mathbf{k}}^{(3)}} \notag \\
	&+&\sum_{m=0}^{5}\frac{\langle \psi_0^{(2)}| c_{\mathbf{k}l'\sigma}^\dagger|\psi_m^{(1)} \rangle \langle \psi_m^{(1)}| c_{\mathbf{k}l\sigma}|\psi_0^{(2)} \rangle}{E_{m,\mathbf{k}}^{(1)}-E_{0,\mathbf{k}}^{(2)}}, \label{eq:G}
\end{eqnarray}
where $| \psi_{m}^{(n)}\rangle$ and $E_{m,\mathbf{k}}^{(n)}$ represent the $m$-th eigenstate and eigenvalue of the $H_{\bf k}^{(n)}$, respectively. 

For $0<U<U_c\approx 6$, the ground state $|\psi_{0}^{(2)}\rangle$ is unique, so that Eq. (\ref{eq:Z2}) can be easily computed and yields  $\nu=1$, indicating a $\mathbb{Z}_2$ topological insulator. However, for $U\geq U_c$, the ground state is four-fold degenerate at the three isolated $M$ points, $(\pm\pi,\pi/\sqrt{3}),(0,2\pi/\sqrt{3})$ (but non-degenerate for all other momenta in the first BZ), posing a technical issue for computing Eq. (\ref{eq:Z2}) since the Green's function is not uniquely defined at these specific momentum points. In fact, there is a level crossing between the ground state and the four degenerate excited states at $U=U_c$, as shown in Fig. (\ref{fig2}a) for $\lambda=0.2$. The state with $E_{0,\mathbf{k}}^{(2)}$ (red line in Fig. (\ref{fig2}a)) involves double occupancy at one of the three sublattices, hence is energetically unfavored at large $U$. While the states with $E_{1,\mathbf{k}}^{(2)}\sim E_{4,\mathbf{k}}^{(2)}$ (green line) do not involve double occupancy, hence stay constant as $U$ increases. At each $M$ point, two of these four states give a parity eigenvalue $\prod_{\mu_\eta>0}\langle \phi_\eta(\mathbf{k})|P_{\mathbf{k}}|\phi_\eta(\mathbf{k}) \rangle=1$ while the other two give $-1$, leading to an ill-defined $\nu$ by directly using Eq. (\ref{eq:Z2}) at $U>U_c$.

Fortunately, the above difficulty can be bypassed by calculating the Pfaffian using Eq. (\ref{eq:Pf}). Figs. (\ref{fig2}b) and (\ref{fig2}c) show the absolute value of Pfaffian $|P(\mathbf{k})|$ within half of the BZ for $U=2$ and $U=8$, respectively. One finds a single zero of $|P(\mathbf{k})|$ at the $K$ point for $U=2$, while four zeros for $U=8$. The odd (even) number of zeros of Pfaffian in half of the BZ suggests a $\mathbb{Z}_2$ nontrivial (trivial) insulator for $U<U_c$ ($U>U_c$).

To support the above conclusion, we further calculate the spin Chern number $C_s$. In our numerical implementation of Eq. (\ref{eq:Cs}), we exclude the three $M$ points with degenerate ground states from the integration grid, as they constitute a measure-zero set and do not contribute to the integral in the thermodynamic limit. Fig. (\ref{fig2}d) shows the calculated $C_s$ as a function of $U$ for $\lambda=0.2$ and different system sizes, where $L=60,90,150,210$ denotes the number of momentum points along each direction of the rhombic BZ. By approaching the thermodynamic limit, our result shows a clear jump of $C_s$ from $-1$ to $2$ at $U=U_c$. The odd and even values of $C_s$ indicate a topological phase transition between $\mathbb{Z}_2$ nontrivial and trivial insulators, consistent with the Pfaffian results.  Moreover, the finite value $C_s=2$ suggests the large-$U$ phase is a nontrivial correlated insulator under the $\mathbb{Z}$ classification \cite{Lin2023,Pixley2024,Lin2024High}. Upon breaking the time-reversal symmetry, this state could potentially undergo a further topological phase transition into a Chern insulator, for example, by spontaneously developing a sufficiently large out-of-plane magnetization \cite{Sheng2011Time,JXYin2022Review}.

Interestingly, the above transition persists to arbitrarily small $\lambda$, with a nearly fixed critical point $U_c\approx 6$. As shown in Fig. (\ref{fig2}e), the spin Chern number is plotted as a function of $\lambda$ for system size $L=210$ and different values of $U$. One observes clearly two topologically distinct phases with $C_s=-1$ and $C_s=2$ even for very small $\lambda$. Fig. (\ref{fig2}f) plots the insulating gap as a function of $U$ for different values of $\lambda$, showing that the spin-orbit coupling induced gap is initially enlarged by the HK interaction, and then gradually suppressed to zero approaching the critical point. Therefore, while weak electron interaction stabilizes the correlated $\mathbb{Z}_2$ topological insulator even for arbitrarily small spin-orbit coupling, strong electron interaction ultimately destroys it. Similar results have been found in previous studies of the Kane-Mele-Hubbard model \cite{Budich2012PRB} and Bernevig-Hughes-Zhang-Hubbard model \cite{Budich2013PRB_BHZ}.

\begin{figure}[t]
	\begin{center}
		\includegraphics[width=8.2cm]{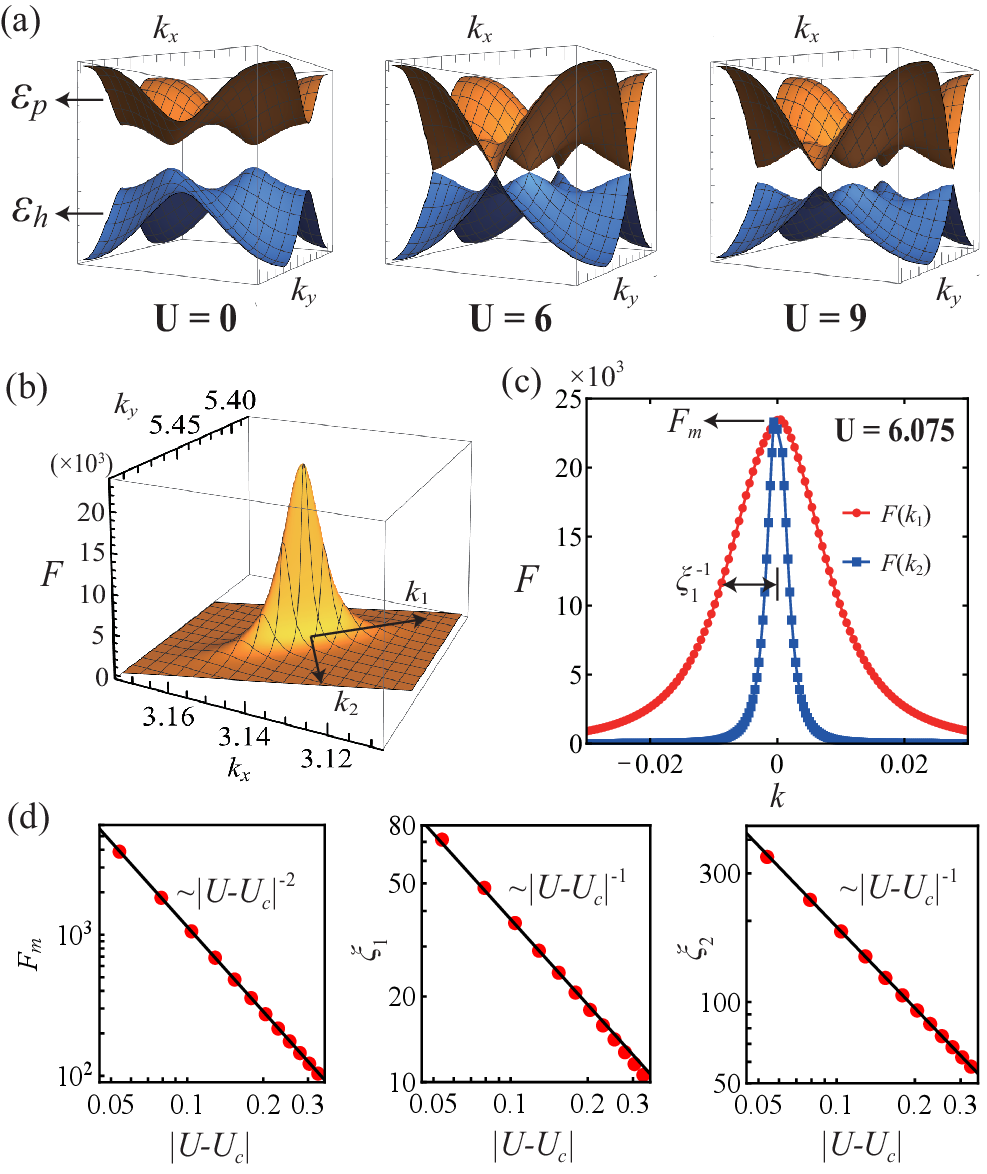}
	\end{center}
	\caption{Critical behaviors of the topological phase transition at $1/3$ filling. (a) The single-particle and single-hole dispersions for $\lambda=0.2$ and different values of $U$. (b) The momentum distribution of Berry curvature at $U=6.075$. (c) The Berry curvature along two different directions ($k_1$ and $k_2$) marked by the black arrows in (b), showing Lorentzian shapes with height $F_m$ and inverse half widths $\xi_1$, $\xi_2$. (d) The scaling behaviors of $F_m$, $\xi_1$ and $\xi_2$ around the critical point.}
	\label{fig3}
\end{figure}

To reveal the nature of topological phase transition at $U_c\approx 6$, we study the evolution of single-particle excitation with increasing interaction.  The particle and hole excitations are defined as $\varepsilon_p=E_{0,\mathbf{k}}^{(3)}-E_{0,\mathbf{k}}^{(2)}$ and $\varepsilon_h=E_{0,\mathbf{k}}^{(2)}-E_{0,\mathbf{k}}^{(1)}$, respectively.  Their dispersion relations are shown in Fig. (\ref{fig3}a) for $\lambda=0.2$ and different values of $U$. One finds that the gap minima shift from the $K$ ($K'$) points to the $M$ points as $U$ increases, and a gap-closing-and-reopening process occurs at the critical point. This is consistent with a generic phase transition between the quantum spin Hall and trivial insulators in the presence of inversion symmetry, in which case the gap can close only at wave vectors $\mathbf{G}/2$, with $\mathbf{G}$ being the reciprocal lattice vectors \cite{Nagaosa2007}. Moreover, since there are three $M$ points in the first Brillouin zone, each one must contribute a jump of spin Chern number $\Delta C_s=1$ at the transition point, consistent with a band inversion of Dirac fermions.   

We also studied the critical behavior around the topological phase transition. It has been demonstrated that the Berry curvature function displays a Lorentzian shape around each gap-closing momentum point, with its height diverging and changing sign at the transition point \cite{Chen2017}. As shown in Figs. (\ref{fig3}b) and (\ref{fig3}c), the Berry curvature around the transition point can indeed be described by the two-dimensional Lorentzian function,
\begin{eqnarray}
	F(\delta \mathbf{k}, U)\approx \frac{F_m}{(1+\xi_1^2\delta k_1^2)(1+\xi_2^2\delta k_2^2)},
\end{eqnarray}  
where $\delta\mathbf{k}=\mathbf{k}-M=k_1\hat{e}_1+k_2\hat{e}_2$, and $F_m$, $\xi_1$ and $\xi_2$ are all functions of $U$. For $M=(-\pi,\pi/\sqrt{3})$, the two orthogonal directions are $\hat{e}_1=(1/2,\sqrt{3}/2)$ and $\hat{e}_2=(\sqrt{3}/2,-1/2)$ (for the other two $M$ points, $\hat{e}_1$ and $\hat{e}_2$ are rotated by $2\pi/3$ and $4\pi/3$, respectively). In the vicinity of the critical point, the three parameters satisfy the scaling laws $F_m\sim \text{sgn}(U-U_c)|U-U_c|^{-\gamma}$, $\xi_{i}\sim |U-U_c|^{\nu_{i}}$ for $i=1,2$. As shown in Fig. (\ref{fig3}d), we find $\gamma=2$, $\nu_1=\nu_2=1$ within numerical accuracy. The relation $\gamma=\nu_1+\nu_2$ is guaranteed by the convergence of the integration of Berry curvature over a small region of area $\sim \xi_1^{-1}\xi_2^{-1}$ around each $M$ point. The above result is consistent with the universality class of topological phase transition of two-dimensional Dirac fermions \cite{Chen2017}.

\subsection{Metal-Mott transition at half filling}

\begin{figure}[t]
	\begin{center}
		\includegraphics[width=8.3cm]{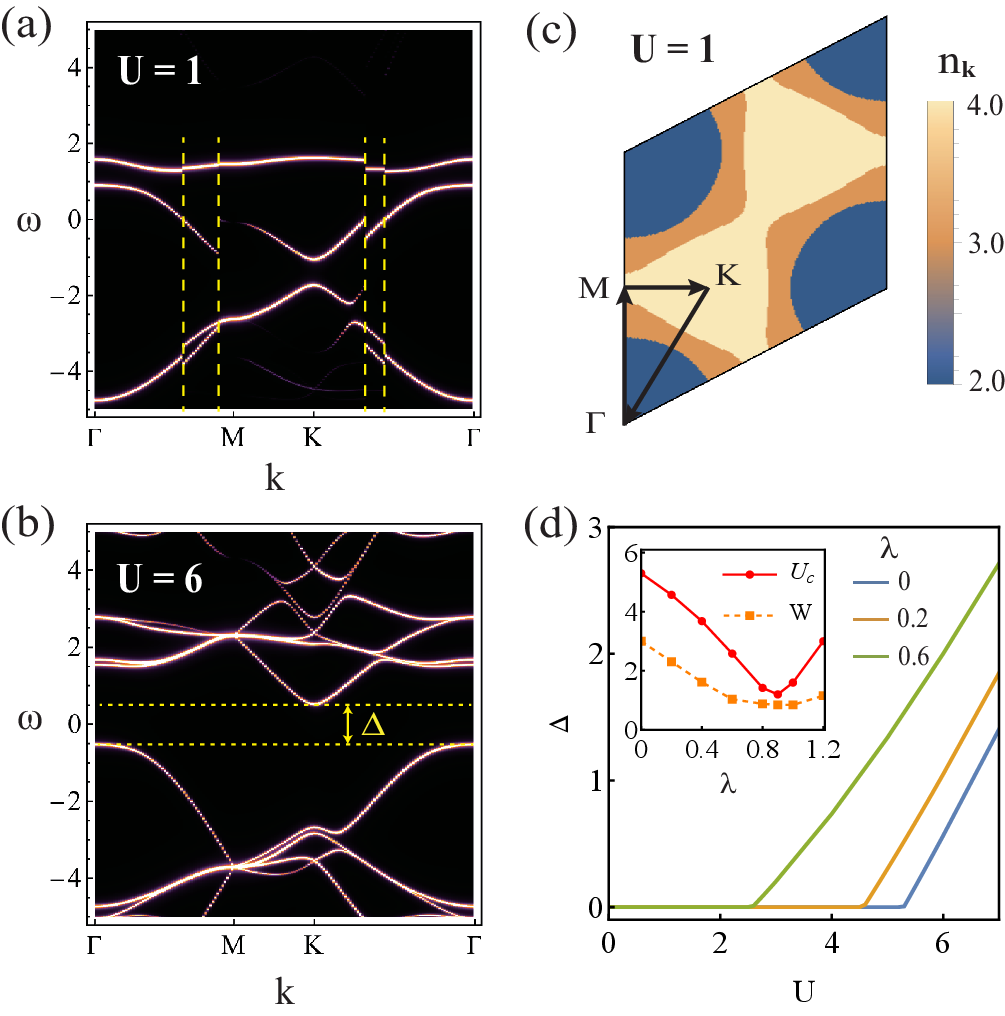}
	\end{center}
	\caption{Single-particle spectra at half-filling for $\lambda=0.2$ and (a) $U=1$, (b) $U=6$. The dashed lines in (a) mark spectra discontinuities at filling surfaces for a non-Fermi liquid metallic phase, while the dotted lines in (b) mark the size of Mott insulating gap. (c) Momentum-resolved electron occupation at $U=1$ and $\lambda=0.2$. (d) The Mott gap $\Delta$ as a function of $U$ for different $\lambda$. The inset shows the critical point $U_c$ and the bandwidth ($W$) of the intermediate band at $U=0$ as functions of $\lambda$. }
	\label{fig4}
\end{figure}

We now turn to the Metal-Mott transition at half filling $n=3$. Fig. (\ref{fig4}a) and (\ref{fig4}b) show the single-particle spectra obtained from the zero-temperature Green's function $G^{11}(\omega,\mathbf{k})$ at $\lambda=0.2$ and two different values of $U$. For the free electron dispersion at $U=0$ (see Fig. (\ref{fig1}b)), the Fermi level cuts the middle band and separates the first BZ into two differently occupied regions: $n_{\mathbf{k}}=2$ and $n_{\mathbf{k}}=4$. A nonzero interaction completely destroys the electron Fermi surface, turning it into a finite region with odd integer filling $n_{\mathbf{k}}=3$, separated from the other two regions with two ``filling surfaces", as shown from the momentum-resolved electron occupation in Fig. (\ref{fig4}c) for $U=1$. The single-particle spectra exhibit discontinuities at the filling surfaces separating differently occupied regions, a typical feature of the HK model \cite{HK1992,Wang2023Kondo,Wang2024HeavyFermion}. Note that the sharp spectra within the $n_{\bf k}=3$ region are not electron-like quasiparticles, since a complete occupation of these bands leads to an odd integer filling. Rather, these are long-lived holon and doublon excitations induced by the local-in-momentum interaction \cite{Phillips2020NatPhys,Zhong2025review}. Therefore, the metallic phase at small $U$ is a non-Fermi liquid.  Further increasing $U$ leads to continuous enlargement of the $n_{\mathbf{k}}=3$ region, which ultimately occupies the entire BZ at a critical point $U_c\approx 4.6$. This gives rise to a Mott insulating state with a nonzero single-particle gap $\Delta$, which separates the upper and lower ``Hubbard bands" consisting of long-lived holons and doublons (see Fig. (\ref{fig4}b) for $U=6$).

Fig. (\ref{fig4}d) shows the size of the Mott gap as a function of $U$ for different spin-orbital coupling strength $\lambda$. As $\lambda$ increases from zero, the critical point $U_c$ first decreases, reaching a minimum $U_c\approx 1.2$ at around $\lambda=0.9$, and then increases with $\lambda$. As shown in the inset of Fig. (\ref{fig4}d), the bandwidth of the middle band at $U=0$ follows the same trend as $\lambda$ increases, suggesting that the Mott transition at half-filling is driven by the competition between kinetic energy and electronic correlation.    

The Mott state at half-filling can be understood from the large $U$ limit, where three electrons at each momentum occupy three sublattices and form three spin-$1/2$ moments, $\boldsymbol{S}_{\mathbf{k}l}=\frac{1}{2}\sum_{\alpha\beta}c_{\mathbf{k}l\alpha}^\dagger \boldsymbol{\sigma}_{\alpha\beta}c_{\mathbf{k}l\beta}$, with the constraint $n_{\mathbf{k}l}=\sum_\sigma c_{\mathbf{k}l\sigma}^\dagger c_{\mathbf{k}l\sigma}=1$.  The second order perturbation of interorbital hopping then leads to an antiferromagnetic exchange coupling between moments at different sublattices (orbitals),  $H_J=\sum_{\mathbf{k} l}J_{\mathbf{k}l}\boldsymbol{S}_{\mathbf{k}l}\cdot \boldsymbol{S}_{\mathbf{k},l+1}$. Here, the coupling strength has a momentum dependence, and is maximized at $\Gamma$ point with $J_{\Gamma}=2(\lambda^2+4t^2)/U$. At each momentum, the three antiferromagnetically coupled spins form a doublet with total spin $S=1/2$.  The many-body ground state is simply a direct product of these $S=1/2$ doublets, which exhibits a large degeneracy similar to that of the single-band HK model \cite{HK1992,Phillips2020NatPhys}. However, here the Mott state shows an internal spin entanglement between different sublattices, a feature beyond the reach of single-band HK models.

\section{Discussion and conclusion}

We briefly discuss other integer filling cases including $n=4$, $n=5$ and $n=1$. For $n=4$, there is also a quantum phase transition from a $\mathbb{Z}_2$ topological insulator to a $\mathbb{Z}_2$ trivial insulator at $U_c\approx 6$. Different from the $n=2$ case, the spin Chern number jumps from $1$ to $-2$ at the transition point, with a quadratic band crossing at each $M$ point.  For $n = 5$,  the system evolves from a non-Fermi liquid metal into a Mott insulator above some critical $U$, in which five electrons per momentum point form a total spin-$1/2$ doublet on one sublattice and fully filled singlets on the other two. Similar Mott state with $n=1$ exists at large $U$, where each momentum point is occupied by one electron with a spin-$1/2$ moment.

In conclusion, we provide an exactly solvable platform on a multi-band frustrated lattice that unifies correlation-driven topological quantum phase transitions and Mott metal-insulator transitions within a single framework. The Kagome lattice HK model thus serves as a potential benchmark for approximate methods applied to interacting topological systems. The continuous topological phase transition with gap closing at the $M$ points demonstrates that electron interaction can drive a weakly-correlated $\mathbb{Z}_2$ topological insulator into a strongly-correlated $\mathbb{Z}_2$ trivial insulator, with a nontrivial spin Chern number. On the other hand, with special care for potential ground-state degeneracy, the Green's function method combined with the momentum-space local nature of the HK interaction provides a useful diagnostic tool for calculating topological invariants in interacting systems.\\

\acknowledgements
We thank Dexi Shao and Chenchao Xu for insightful discussions. This work was supported by the National Natural Science Foundation of China (Grant No. 12304174).

\end{document}